\documentclass[notitlepage,pre,proof, twocolumn,longbibliography, floatfix]{revtex4-1}%
\usepackage{amssymb}
\usepackage{amsfonts}
\usepackage{amsmath}
\usepackage{graphicx}%
\providecommand{\U}[1]{\protect\rule{.1in}{.1in}}
\begin{document}
\title{Long-wavelength density fluctuations as nucleation precursors}
\author{James F. Lutsko}
\homepage{http://www.lutsko.com}
\email{jlutsko@ulb.ac.be}
\author{Julien Lam}
\affiliation{Center for Nonlinear Phenomena and Complex Systems CP 231, Universit\'{e}
Libre de Bruxelles, Blvd. du Triomphe, 1050 Brussels, Belgium}

\begin{abstract}
Recent theories of nucleation that go beyond Classical Nucleation Theory
predict that diffusion-limited nucleation of both liquid droplets and of
crystals from a low-density vapor (or weak solution) begins with long-wavelength density
fluctuations. This means that in the early stages of nucleation, ``clusters''
can have low density but large spatial extent, which is at odds with the
classical picture of arbitrarily small clusters of the condensed phase. We
present the results of kinetic Monte Carlo simulations using Forward Flux Sampling to show that these
predictions are confirmed: namely that on average, nucleation begins in the
presence of low-amplitude, but spatially extended density fluctuations thus
confirming a significant prediction of the non-classical theory.

\end{abstract}
\date{\today }
\maketitle

\section{Introduction}

Nucleation is a widely studied phenomenon that plays
an important role in such diverse subjects as the function of
cells\cite{cells}, stellar evolution\cite{white_dwarfs} and
superhydrophobicity\cite{superhydro}. The particular example of
crystallization is of great importance in chemistry, materials science and
fundamental physics and has been found to be far more complex than once
believed. For this reason, the study of nucleation pathways - the steps by
which thermal fluctuations build up a critical cluster - has become a major
focus of modern work by experimentalists, \ simulators and theorists \ alike.

In Classical Nucleation Theory\cite{Kashchiev} (CNT) the nucleation of new phases is assumed to begin with the formation of small clusters that then evolve via the stochastic addition and loss of material. For example, in the most naive version, the formation of a crystal is supposed to begin with small oligomers of a few growth units arranged the final crystalline form. This crystallite then grows while retaining its ordered structure. Droplets nucleating from a vapor and the converse, bubbles nucleating from a liquid, are imagined to proceed in analogous manners. The idea that some sort of precursor process could play a role is one that has been suggested in several contexts. For example, Shen and Debenedetti\cite{Shen} reported on Monte Carlo simulations showing large ramified voids as precursors to bubble nucleation in superheated fluids. These results were challenged by Wang, Valeriani and Frenkel\cite{Wang} who used dynamical simulations and who reported compact subcritical structures. They attribute the difference in the results to the choice of order parameter (local versus global). Intriguingly, these authors did report on another type of precursor: namely, that bubble nucleation seemed to preferentially begin in ``hotspots'' of locally elevated temperature.

There has long been speculation that local concentration would play a role in the nucleation of condensed phases such as a droplet from a liquid. In the 1960's, Russell\cite{Russell} coupled the usual Becker-Doring dynamics of CNT to simple models of the surrounding fluid and concluded that critical nuclei were likely to occur in enriched areas of the system. As Russell notes, when material clumps together to form a dense cluster, the surrounding system is necessarily depleted unless new material flows in to replace it. In his pioneering work on field theoretic approaches to nucleation, Langer\cite{Langer, Langer2} also discusses the importance of coupling cluster formation to the local concentration but his work was mostly concerned with determining the nucleation rate and so focussed on the endpoint of nucleation - the critical cluster. More recently, Peters\cite{Peters} discussed a more detailed model of the coupling between the local concentration and cluster dynamics and concluded that local concentration gradients play a key role in determining the fate of clusters. 

One approach that fully couples the fluctuations leading to cluster formation and transport processes is Mesoscopic Nucleation Theory (MeNT)
which combines classical Density Functional Theory methods with
fluctuating hydrodynamics and techniques from stochastic process
and large deviation theory to give a complete, unconstrained framework for
studying nonclassical pathways\cite{Lutsko_JCP_2011_Com,Lutsko_JCP_2012_1,
Lutsko_HCF}.\ Recent successes of this approach has been the establishment of
the connection between it and CNT\cite{Lutsko_NJP}, demonstrating the latter
as a particularly simple approximation to the former, and the unbiased
description of pathways for crystallization\cite{Lutsko_HCF}.\ 

\begin{figure}
[t]\includegraphics[angle=0,scale=0.25]{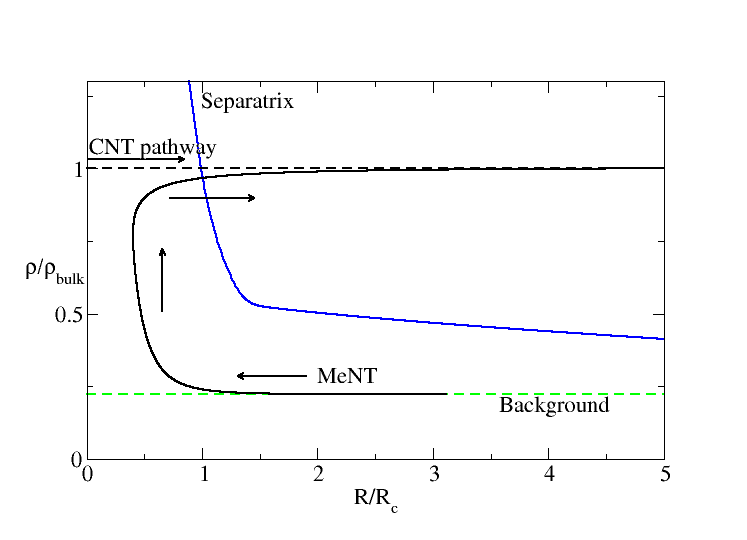}

\caption{Comparison of nucleation pathways predicted by MeNT (full black
line)  and CNT (broked back line) for the formation of a dense
(concentrated) phase from a weak solution. The evolving nucleus is
characterized by its interior concentration $\rho$  (expressed as a ratio to
the bulk density of the condensed phase) and by its radius $R$ (expressed
relative to the critical radius). The broken horizontal line at low
concentration is the background, or initial, concentration of the weak
solution.  In CNT, the interior density is equal to that of the bulk and is
constant and only the radius changes whereas in MeNT both variables change
continuously and the path begins with a small density excess over a
substantial area. Also shown are the separatrix which must be crossed to
achieve nucleation: the critical cluster for each pathway is the
intersection between the pathway and the separatrix.} \label{fig0}
\end{figure}

The main application of MeNT has so far been to diffusion-limited nucleation
such as for liquid-liquid phase separation or crystallization of large
molecules in solution. One key difference in the predicted pathways in this
case, with respect to CNT, but consistent with the earlier work mentioned, is that they always begin with a long-wavelength,
low-amplitude density fluctuation\cite{Lutsko_JCP_2011_Com,Lutsko_JCP_2012_1}%
.\ These can be thought of as clusters which are large in spatial extent but
have a density only slightly greater than the background (see Fig. \ref{fig0})
or, alternatively, not as a single structure but simply a region of the system
with a slightly elevated density. Such regions always exist due to thermal
fluctuations: for example, given a snapshot of any simulation and dividing the
volume in two, one will always find the density of one half slightly above the
system average and the other slightly below.\ The theory predicts that
nucleation of a condensed phase is more likely to occur in the former than in the
latter which makes intuitive sense.\ This result has proven very robust being
found both for nucleation of liquid droplets from a vapor and for that of
crystals from a weak solution and by means of simple, coarse-grained theories
derived from MeNT\cite{lutsko2014a} as well as in large-scale calculations
based on the full theory\cite{Lutsko_HCF}.\ It also persists for both the use
of simple squared-gradient free energy models as well as for sophisticated
DFT's.\ Ultimately, it is a result of the dynamics correctly conserving mass.
In this work, we describe a validation of this prediction based on dynamical simulations.

In this work, the goal is to verify this element of the nucleation pathway via simulation. This goal is difficult as one cannot use the usual definitions of a ``cluster''. For example, if one defines a cluster to be a collection of molecules that are each within some fixed distance of another member of the cluster, then one is biasing the definition to regions with a local density exceeding some threshhold. This automatically rules out observation of a ``cluster'' consisting of a slight increase in density above the background.  For this reason, we devote the next Section to a detailed description of our simulation strategy. We describe a novel  use of Forward Flux Sampling to determine the nucleation pathway for an open system which allows us to identify a local increase in density. We present our simulation results in the following Section where it is shown that nucleation is most likely to begin with a long-wavelength, small amplitude density fluctuation. We end with a short summary of our Conclusions.  
 
\section{Simulation Strategy}

Our strategy is inspired by recent calculations showing
that the long-wavelength fluctuations manifest themselves as an overall
increase in density in open, finite systems prior to cluster
formation\cite{Lutsko_HCF}. In order to simulate three-dimensional systems
which are both open and yet with a mass-conserving dynamics, we use periodic
boundaries in two directions and allow molecules to move out of the system in
the natural course of the dynamics, with this being balanced by molecules
randomly entering through the same boundaries. The number of molecules within
the simulation cell therefore fluctuates while in the interior of the cell,
mass is strictly, locally conserved. This model corresponds to  a subvolume of a
large system.

We have perfomed kinetic Monte Carlo simulations for molecules on a lattice
with lattice constant $a$ and with these semi-open boundary conditions. Our
simulations include a single species of molecules that bonds to any nearest neighbors in
the six Cartesian directions with an energy $-\epsilon$ for each bond. (We scale all the temperature and all energies to this quantity so its physical value is not needed.)
Molecules jump from one lattice site to an unoccupied nearest-neighbor lattice
site with a rate proportional to $\min(1,\exp(-\beta\Delta E))$ where $\Delta
E$ is the total change in the system energy between the final and inital
states, $\beta= 1/k_{B}T$, $k_{B}$ is Boltzmann's constant and $T$ is the
temperature. In the low-density limit, this reduces to a three-dimensional
random walk and hence, macroscopically, to diffusion. Note that only monomers
move: there is no provision for the collective movement of clusters. This is
important because it means that only monomers can attach to clusters and the
concentration of monomers is generally less than the total number of molecules
in the cell.

\begin{figure}
[t]\includegraphics[angle=0,scale=0.35]{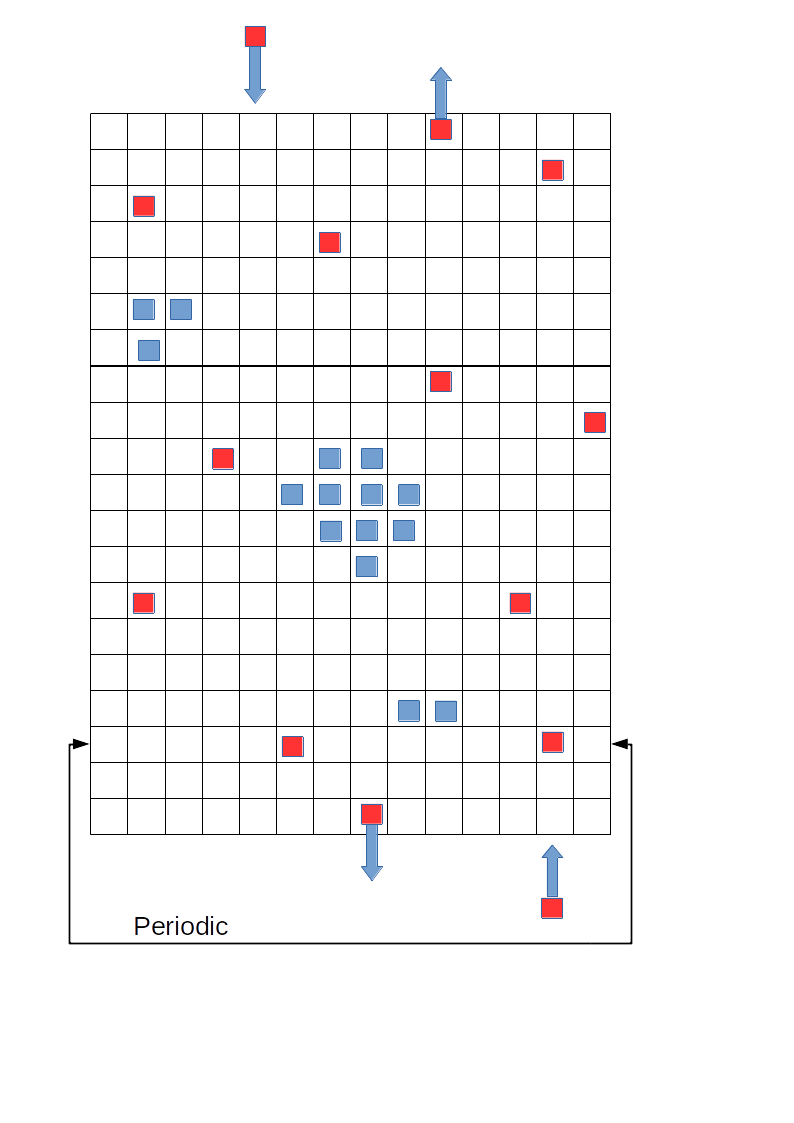}

\caption{Schematic illustation of the simulation scheme projectd onto two dimensions. The simulations take place on a grid with molecules randomly hopping between nearest neighbor sites. The boundaries are periodic in two dimensions and open in the third: molecules that jump out of the simulation via the top and bottom boundaries are lost to the simulation but balanced by molecules randomly entering at rates determined by the applied chemical potential. Clusters of molecules - i.e. those connected by nearest neighbor bonds -  are shown as blue.}
\label{grid}
\end{figure}

The control parameters in the simulation are the temperature and the number of
molecules entering per time step via the open surfaces which, in turn,
controls the average number of molecules in the simulation cell and the number
of monomers. In equilibrium, with an average density of monomers $n_{1} $, the
number of monomers exiting through the boundaries in each time step will be
$2n_{1}Aa \nu_{0} \delta t$ where $A$ is the area of the open boundaries and
$\nu_{0}$ is the ``attempt frequency''. (The latter is the fundamental time scale of the simulations: all times are expressed in terms of it and so its actual value is arbitrary.) If the number of monomers entering per
time step, $\delta t$, is $2N_{B}$, then the concentration in equilibrium is
given by $n_{1}Aa \nu_{0} \delta t = N_{B}$. Balancing the rates of attachment
and detachment at a kink site similarly show that at two-phase coexistence the
concentration of monomers is $n_{1}^{\text{coex}} = e^{-3\beta\epsilon}a^{-3}$\cite{Lutsko_kink}
so that $N_{B}^{\text{coex}} = e^{-3\beta\epsilon}Aa^{-2} \nu_{0} \delta t$. When presenting results below, we use $S \equiv ln \left(N_{B}/N_{B}^{\text{coex}}\right)$ as a measure of the supersaturation. The simulation scheme is illustrated in Fig. \ref{grid}.

For the simple lattice-based kMC dynamics there are only two phases: a
low-density vapor and a crystalline state. Crystalline clusters are identified
via the nearest neighbor bonds giving the cluster size as the natural order
parameter; specifically, for any configuration of positions, the order
parameter is taken to be the size of the largest cluster present in the
system. For the conditions considered here, a system beginning in the vapor
state will seldom spontaneously form clusters larger than a few molecules and
those are, in general, short-lived. Thus we use a standard rare-event
technique, Forward Flux Sampling\cite{FFS1,FFS2} (FFS), to drive the system in
an unbiased manner from the initial, low-density state to crystallization.
The key concept in FFS\ is a set of surfaces in phase space defined as being
all configurations with a given value of the order parameter. In our
implementation of FFS for this system, we begin with the vapor and allow the
system to evolve.\ Each time a configuration having largest cluster size
$N_{0}-1$ evolves into one having largest cluster size $N_{0}$ in the next
time timestep, we store the latter configuration thus creating a database,
$D_{N_{0}},$ of configurations of systems crossing the $N_{0}$ surface in the
direction of increasing cluster size. Once this database has a given number of
configurations, the second part of the simulation begins. This consists of
randomly choosing one of the configurations from $D_{N_{0}}$ and evolving the
sytem forward in time. If it eventually returns to the initial metastable
basin, i.e. the vapor state having order parameter less than a specified value
$N_{\text{equil}}$, then the simulation is discarded and the process is
repeated. If, before returning to the initial basin, the system evolves into
one with order parameter $N_{0}+1$, the simulation is stopped, the
configuration recorded in a new database, $D_{N_{0}+1}$, and the process is
repeated until $D_{N_{0}+1}$ has the required number of configurations. Once
complete, this continues for order parameter $N_{0}+2$, etc until some maximal
value $N_{\text{max}}$ is reached. In the initial stages, most trajectories
fail to reach the next surface but as the cluster sizes increase, the success
rate does as well until almost all trajectories successfully reach the next
surface. Thus, the configurations in $D_{N_{\text{max}}}$ are effectively
stable crystals that will only grow with time.

From the empirically determined probabilities of trajectories launched from
one surface to reach the next, and from the empirically determined rate at
which trajectories leave the initial basin, the transition rate can be
determined. Here, however, the main interest is not the rate but the paths.
For any given element of $D_{N}$ one can trace back the parent configuration
in $D_{N-1}$ and the grandparent in $D_{N-2}$ etc and in this way the entire
trajectory going back to an initial configuration in $D_{N_{0}}$. If we denote
by $\Gamma_{N}^{\left(  j\right)  }$ the $j$-th configuration in database
$D_{N}$, then a successful trajectory leading to a stable crystal will be a
sequence of snapshots $\left(  \Gamma_{N_{0}}^{\left(  j_{0}\right)  }%
,\Gamma_{N_{0}+1}^{\left(  j_{1}\right)  },...,\Gamma_{N_{\text{max}}%
}^{\left(  j_{N_{\text{max}}-N_{0}}\right)  }\right)  .$Any configuration
$\Gamma_{N}^{\left(  j\right)  }$ will be a member of some number, $n^{\left(
S\right)  }\left(  \Gamma_{N}^{\left(  j\right)  }\right)  $, of such
successful trajectories. If we also record for each configuration the total
number of times it was used as an initial value, $n\left(  \Gamma_{N}^{\left(
j\right)  }\right)  $, then we also know the total number of unsuccessful
trajectories launched from this configuration, $n^{\left(  U\right)  }\left(
\Gamma_{N_{0}}^{\left(  j\right)  }\right)  =n\left(  \Gamma_{N}^{\left(
j\right)  }\right)  -n^{\left(  S\right)  }\left(  \Gamma_{N}^{\left(
j\right)  }\right)  $. This information is used to compute three types of
average of the mass. Let $M\left(  \Gamma_{N}^{\left(  j\right)  }\right)  $
be the mass, or number of molecules, in configuration $\Gamma_{N}^{\left(
j\right)  }$. Then, first, we define the simple average mass of a database, that is, the average mass of all configurations possessing a cluster of given size $N$, 
\begin{equation}
\left\langle M\right\rangle _{N}\equiv\frac{\sum_{\Gamma_{N}^{\left(
j\right)  }\in D_{N}}M\left(  \Gamma_{N}^{\left(  j\right)  }\right)  }%
{\sum_{\Gamma_{N}^{\left(  j\right)  }\in D_{N}}1}.
\end{equation}
Next is a \emph{weighted-average} number of molecules in each database where the
weight for each configuration $\Gamma_{N}^{\left(  j\right)  }\in D_{N}$ is the
number of successful trajectories that include it,
\begin{equation}
\left\langle M\right\rangle _{N}^{\left(  S_{1}\right)  }\equiv\frac
{\sum_{\Gamma_{N}^{\left(  j\right)  }\in D_{N}}n^{\left(  S\right)  }\left(
\Gamma_{N}^{\left(  j\right)  }\right)  M\left(  \Gamma_{N}^{\left(  j\right)
}\right)  }{\sum_{\Gamma_{N}^{\left(  j\right)  }\in D_{N}}n^{\left(
S\right)  }\left(  \Gamma_{N}^{\left(  j\right)  }\right) }%
\end{equation}
We have also defined a variant called $\left\langle M\right\rangle
_{N}^{\left(  S_{2}\right)  }$ by replacing $n^{S}\left( \Gamma_{N}^{\left(
j\right)  }\right) $ as the weight by $\frac{n^{S}\left( \Gamma_{N}^{\left(
j\right)  }\right) }{n\left( \Gamma_{N}^{\left(  j\right)  }\right) }$ (for
the cases that $n\left( \Gamma_{N}^{\left(  j\right) }\right) >0$) to thus
correct for the possibility that a configuration is chosen many times and only
yields a small fraction of successful trajectories. Finally, we define the
average mass of completely unsuccessful configurations: that is, those
configurations that were used as initial values but did not lead to any
successful trajectories,
\begin{equation}
\left\langle M\right\rangle _{N}^{\left(  U\right)  }\equiv\frac
{\sum_{\substack{\Gamma_{N}^{\left(  j\right)  }\in D_{N} \\n\left(
\Gamma_{N}^{\left(  j\right)  }\right)  >0}}\delta_{n^{\left(  S\right)
}\left(  \Gamma_{N}^{\left(  j\right)  }\right)  =0}M\left(  \Gamma
_{N}^{\left(  j\right)  }\right)  }{\sum_{\substack{\Gamma_{N}^{\left(
j\right)  }\in D_{N} \\n\left(  \Gamma_{N}^{\left(  j\right)  }\right)
>0}}\delta_{n^{\left(  S\right)  }\left(  \Gamma_{N}^{\left(  j\right)
}\right)  =0}}.
\end{equation}

\section{Results}

\begin{figure}[t]\includegraphics[angle=0,scale=0.35]{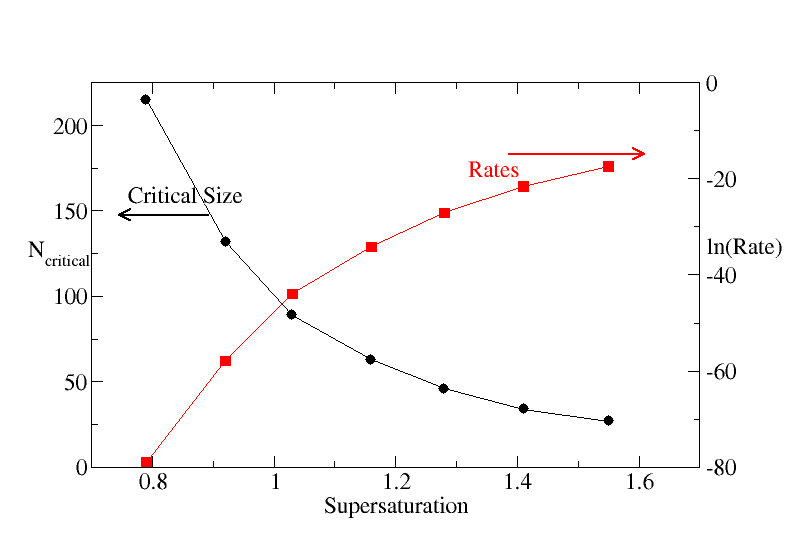}
  \caption{The number of molecules in the critical cluster (circles, left-hand scale) and the log of the measured nucleation rates (squares, right-hand scale) as functions of the supersaturation at $k_{B}T=0.4\epsilon$. The nucleation rate is for the entire volume of the simulation cell and is expressed in terms of the fundamental KMC timescale defined by $\nu_{0}$.}
\label{fig01}
\end{figure}

We have performed simulations on a system with simulation cell of size $50
\times50 \times50$, a temperature of $k_{B}T=0.4 \epsilon$. The FFS calculations were performed using 
approximately $160,000$ configurations in each database. (This number was chosen simply empirically as one in which the noise in the final results was sufficiently small to reveal the trends.) At these conditions, the equilibrium system has a concentration of approximately $e^{-3/k_{B}T}=5.5 \times 10^{-4}$ monomers per unit cell or $69$ monomers on average in the whole simulation cell. Our nonequilibrium simulations spanned the range of supersaturations from $S=0.79$ (about $170$ molecules) to $S=1.55$ (about $394$ molecules).  
To give some context to our analysis of the nucleation pathways, we show in Fig. \ref{fig01} the critical clusters and nucleation rates as functions of the supersaturation. The critical cluster contains about $215$ molecules at the lowest supersaturation down to only $27$ at the highest concentrations. Note that the critical clusters are easily determined from the FFS simulations since they directly provide the probability of a cluster of size $N$ growing to size $N+1$ and that one of size $N+1$ grows to size $N+2$ etc.: mulitplying these gives the probability that a cluster of size $N$ grows indefinitely and the cluster for which this probability is $50\%$ is the critical cluster.  

\begin{figure}[t]\includegraphics[angle=0,scale=0.35]{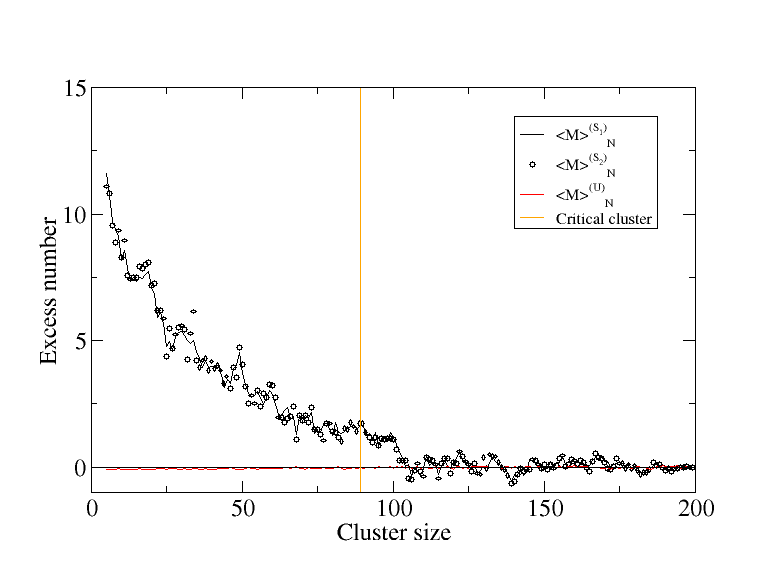}

\caption{Difference between the average number of molecules as computed by
using the various weightings and the reference, unweighted average, as a
function of cluster size. The vertical line indicates the critical cluster
which contains approximately $90$ molecules (supersaturation = $1.03$) and the average number of
monomers in the simulation cell in the initial metastable-state  was approximately $220$.}
\label{fig1}
\end{figure}

Figure \ref{fig1} shows the difference between the  weighted-average number of molecules and the unweighted average for the three types of weighted average as functions of the cluster size for the case $S=1.03$ as determined from the FFS simulations. It is clear that both
the $S_{1}$ and $S_{2}$ weightings give virtually identical results and that
both show a systematic excess relative to the unweighted average during the
nucleation phase. The average over unsuccessful configurations, however, shows
only a very slight decrease relative to the reference. These observations were
the same in all other simulations reported below.

\begin{figure*}[tbh]
\begin{center}
\includegraphics[angle=0,scale=0.35]{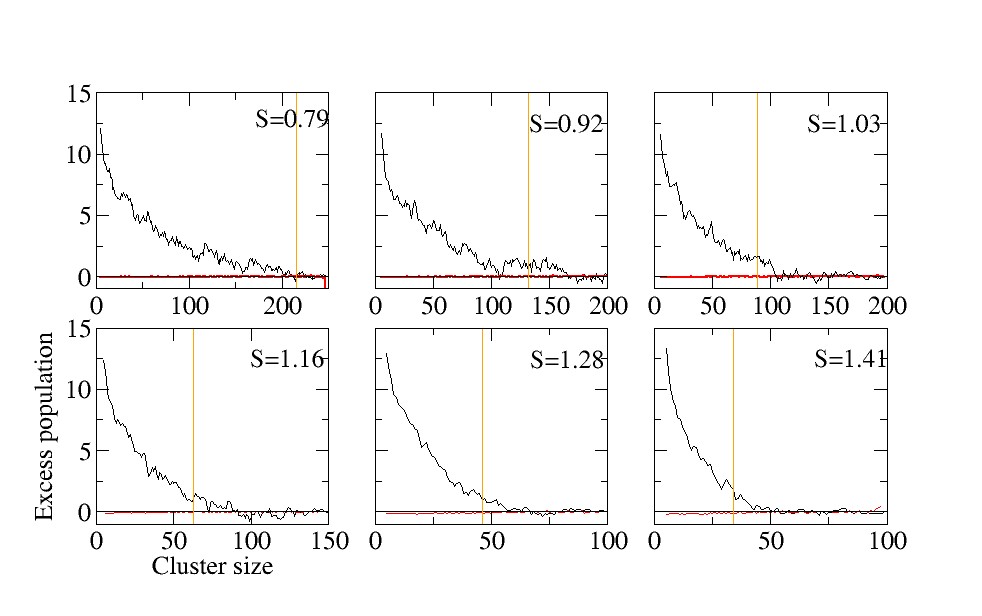}
\end{center}
\caption{Difference between the average number of molecules in the simulation
cell calculated by weighting each configuration by the number of successful
trajectories it participates in and the average overall all configurations as
a function of the size of the clusters. The different panels are for different
supersaturations and in each case the critical cluster is indicated by a
vertical line. The temperature was $k_{B}T=0.4 \epsilon$ and the simulation
cell was a cube with $50$ sites per side. Each figure is labeled by the
supersaturation $S=<N>/<N>_{\text{coex}}$ where $<N>$ is the average number of
molecules in the initial state and $<N>_{\text{coex}}$ is that for
coexistence. }%
\label{fig2}%
\end{figure*}

Figure \ref{fig2} shows the excess number of molecules $\Delta N$ in the
successful trajectories relative to the average over all configurations at
each value of $N$ for different supersaturations corresponding to critical
clusters ranging from 35 to 220 molecules. It is clear that there is a
systematic trend towards excess numbers of molecules in the successful
trajectories thus confirming the theoretical prediction. For the smallest
clusters, the excess amounts to approximately $5\%$ of the average number of
molecules in equilibrium.

\begin{figure}
[htb]\includegraphics[angle=0,scale=0.25]{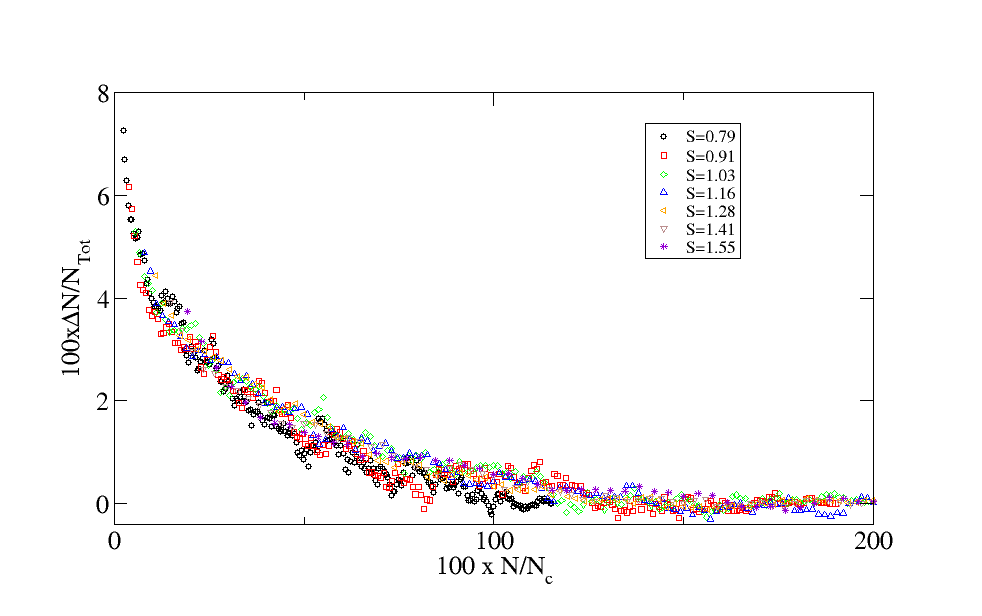}
\caption{The data from Fig. \ref{fig2} but with the number of molecules in
the clusters scaled by the critical size and the excess number of molecules
scaled by the total number in the simulation cell in equilibrium. }
\label{fig3}
\end{figure}

We also show in Fig. \ref{fig3} the same data with $\Delta N$ scaled to the
average number of molecules in the simulation cell in equilibrium and with $N
$ scaled to the critical radius. The data collapse to a single curve thus
further supporting the systematic nature of the observed excess in the
successful trajectories.

Our  interpretation of these results is that in the early stages of nucleation, the total number of molecules in the simulations varies stochastically and that configurations with a higher than average density are more likely to participate in successful nucleation events. This is in agreement with the predictions of MeNT and serve to support its more  complex description of nucleation, compared to CNT. 

\section{Conclusions}

We have used simple kMC simulations to confirm a
robust prediction of MeNT: namely, that diffusion-limited nucleation is most
likely to begin with a long-wavelength, small amplitude density fluctuation.
This supports the nuanced picture of the process that arises when coupling mass
transport, mass conservation and thermodynamics.The results do not themselves
invalidate CNT - at least at low supersaturations, the difference between the
nucleation pathways predicted by CNT and observed in the simulations is only
significant in the early stages of nucleation - but they do serve to show that even in
relatively simple cases, the intuitively appealing classical picture has
limitations.One practical consequence is that studies that rely on local
density to define clusters may miss the complexity of the early stages of
nucleation completely. Indeed, referring back to Fig. \ref{fig0}, note that if one imposes a minimum density so that anything below that is not identified as a ``cluster'', then observed clusters will have a  minimum size that is larger than zero. This is consistent with previous work  of Trudu, Donadio and Parrinello\cite{Trudu}  who reported such an observation in simulations of crystal nucleation: namely, that precritical crystallites occur only above some finite size. 
It also suggests that the manipulation of density
fluctuations, e.g. as reported in the GradFlex microgravity
experiment\cite{gradflex}, could be useful in controlling nucleation.

Finally, we speculate that while our present results are applicable to a particular form of nucleation - namely, that of a dense phase from a dilute phase - analogous phenomena could well occur in other systems. For example, density fluctuations can be both positive and negative: one might expect bubbles to preferentially nucleate in a liquid in regions of lower-density fluctuations. When liquids freeze into solids or, vice versa, solids melt to form liquids the change in density is often relatively small  but in these cases, heat transport is important due to the latent heat involved in the phase transitions and so long-wavelength, small amplitude temperature fluctuations could be the analog to the precursors studied here - i.e., the ``hot-spots'' and ``cold-spots'' alluded to in the Introduction.  
\bigskip

\begin{acknowledgments}
This work of JFL was supported by the European Space Agency (ESA) and the Belgian
Federal Science Policy Office (BELSPO) in the framework of the PRODEX
Programme, contract number ESA AO-2004-070. JL acknowledges financial support of the Fonds de la
Recherche Scientifique - FNRS. Computational resources have been
provided by the Shared ICT Services Centre, Universit Libre de Bruxelles
\end{acknowledgments}

%

\end{document}